 \documentclass[aps,prl,reprint,preprintnumbers,showpacs,showkeys,superscriptaddress,floatfix]{revtex4-1}
\usepackage{graphicx}
\usepackage{epsfig,amssymb,amsmath,color}
\usepackage{epstopdf}
\usepackage{times}
\usepackage[colorlinks=true,citecolor=blue,linkcolor=blue]{hyperref}

\usepackage{ulem}

\newcommand{\beq}{\begin{equation}}
\newcommand{\eeq}{\end{equation}}
\newcommand{\bea}{\begin{eqnarray}}
\newcommand{\eea}{\end{eqnarray}}
\newcommand{\bear}{\begin{array}}
\newcommand {\eear}{\end{array}}
\newcommand{\bef}{\begin{figure}}
\newcommand {\eef}{\end{figure}}
\newcommand{\bec}{\begin{center}}
\newcommand {\eec}{\end{center}}

\def\lrfp#1#2#3{ \left(\frac{#1}{#2} \right)^{#3}}

\begin{document}
\title{Domain Wall Formation from Level Crossing  in the Axiverse}

\author{Ryuji Daido}
\email{daido@tuhep.phys.tohoku.ac.jp}
\affiliation{Department of Physics, Tohoku University, Sendai 980-8578, Japan}

\author{Naoya Kitajima}
\email{kitajima@tuhep.phys.tohoku.ac.jp}
\affiliation{Department of Physics, Tohoku University, Sendai 980-8578, Japan}

\author{Fuminobu Takahashi}
\email{fumi@tuhep.phys.tohoku.ac.jp}
\affiliation{Department of Physics, Tohoku University, Sendai 980-8578, Japan}
\affiliation{Kavli Institute for the Physics and Mathematics of the
  Universe (WPI), Todai Institutes for Advanced Study, University of Tokyo,
  Kashiwa 277-8583, Japan}

\begin{abstract}
We point out that domain wall formation is a more common phenomenon in the Axiverse
than previously thought.  Level crossing could take place if there is a mixing between axions,
and if  some of the axions acquire a non-zero mass through non-perturbative effects as
the corresponding gauge interactions become strong.   The axion potential
changes significantly during the level crossing, which affects the axion dynamics in various ways. 
We find that, if there is a mild hierarchy in the decay constants, the axion starts to run along the valley of the 
potential, passing through many crests and troughs,  until it gets trapped in one of the minima; the
{\it axion roulette}. The axion dynamics exhibits a chaotic behavior during the oscillations,
and which minimum the axion is finally stabilized  is highly sensitive to the initial misalignment angle.
Therefore, the axion roulette is considered to be accompanied by domain wall formation.
The cosmological domain wall problem can be avoided by introducing a small bias between the vacua.
We discuss cosmological implications of the domain wall annihilation for baryogenesis and 
future gravitational wave experiments.

\end{abstract}
\preprint{TU-995, IPMU15-0076}
\maketitle

\paragraph{Introduction}
-- There may be many axions in nature.  In string theory, compactifications of the extra
dimensions often offer a large number of axions~\cite{Svrcek:2006yi,Blumenhagen:2006ci}.
The Universe with a plenitude of such axions whose masses range over many orders of magnitude
is called the Axiverse~\cite{Arvanitaki:2009fg}.
The striking feature of the axions is that they respect shift symmetry, which is unbroken at the perturbative
level. Non-perturbative effects, however,  typically break the shift symmetry to a discrete one, generating a periodic 
potential for the axions. Some of the axions may remain relatively light and play an important role in cosmology
such as inflation~\cite{Freese:1990rb,Czerny:2014wza}, dark energy~\cite{Kim:2009cp,Arvanitaki:2009fg},
dark matter~\cite{Ringwald:2012hr,Kawasaki:2013ae},  and baryogenesis~\cite{Cohen:1987vi,Chiba:2003vp,Kusenko:2014uta,Daido:2015gqa}.

The axion mass is not necessarily a constant, but can vary with time.
For example, the QCD axion acquires
a non-zero mass at the QCD phase transition, while it is
massless at high temperature. Similarly, some of the axions may
acquire a mass after inflation. This is expected to be the case if the 
the hidden sector is heated to high temperatures where
the hidden gauge coupling is weak.
As the temperature of the hidden sector decreases in the course of the cosmic expansion,
the axion mass is generated when the hidden gauge interactions become non-perturbative. 

In general, axions have mass and kinetic mixings with each 
other~\footnote{Such mass and kinetic mixings among multiple axions are exploited recently in the context of 
inflation~\cite{Kim:2004rp,Choi:2014rja,Higaki:2014pja,Bachlechner:2014hsa}.}.
The level crossing takes place if the mass of one (or more) of the axions increases with time
due to the non-perturbative effects and becomes heavier than  other axions.
During the level crossing, the axion potential changes significantly, which affects the axion dynamics in 
various ways. The level crossing phenomenon among the axions in the Axiverse
has not attracted attention so far to the best of our knowledge, and we shall study its cosmological impact
in this Letter.

The axion starts to oscillate when its temperature-dependent mass becomes comparable to
the Hubble parameter.  If the axion starts to oscillate well before the level crossing, the change of the
axion potential is sufficiently slow compared to the oscillation period. In this case, we find that the resonant transition takes place at the level crossing
 {\it a la} the MSW effect in neutrino physics~\cite{Wolfenstein:1977ue,Mikheev:1986gs}.
On the other hand, if the commencement of the axion oscillations is close to the level crossing, 
the axion does not really oscillate along one direction, but starts to run along the valley of the potential, passing through
many crests and troughs, until it gets trapped in one of the potential minima.   
We find that which minimum the axion is finally stabilized is
highly sensitive to the initial misalignment angle, and as we shall see, its dependence is rather chaotic.
We call the chaotic run-away behavior  the ``{\it axion roulette}."

Because of the high sensitivity to
the initial position and possible spatial instabilities,  the axion roulette is expected to be accompanied 
by domain wall formation. Therefore, domain wall formation might be a more common phenomenon in the
Axiverse than previously thought.  We will determine the conditions for the axion roulette to take place.
The domain walls are cosmologically problematic, and they must be either inflated away, or unstable and decay rapidly.
We shall discuss the implications of the domain wall collapse for baryogenesis and future gravitational wave experiments.

Lastly let us briefly mention the works in the past. The resonant transition between the QCD axion
and the hidden axion was studied in the pioneering paper by Hill and Ross~\cite{Hill:1988bu}. 
Recently two of the present authors (NK and FT) studied the resonant transition in Ref.~\cite{Kitajima:2014xla}, 
where the adiabatic invariant was correctly identified, and it was shown that not only the axion abundance 
but also its isocurvature perturbations can be significantly suppressed if the adiabaticity is weakly broken 
by the initial condition of the axion close to the hilltop. As emphasized in Ref.~\cite{Kitajima:2014xla},
the level crossing phenomenon is not limited to the QCD  axion and the hidden axion, but it occurs commonly 
between any axions with a mixing.  The purpose of this Letter is to show that 
the level crossing leads to not only resonant transition, but also domain wall formation.

\paragraph{Level crossing}
-- Let us consider the low-energy effective Lagrangian for axions $a_1$ and $a_2$;
\beq
	\mathcal{L} = \sum_{i=1,2} \frac{1}{2} \partial^\mu a_i \partial_\mu a_i - 
	V_1(a_1,a_2)  - V_2(a_1,a_2)
	\label{Va1}
\eeq
with
\bea
	\label{Va2}
V_1(a_1,a_2)&=& \Lambda_1^4 \left(1-\cos \left(n_1\frac{a_1}{f_1}+n_2\frac{a_2}{f_2}\right) \right) \\[1mm]
V_2(a_1,a_2) &=& m_a(T)^2 f_2^2 \left(1-\cos \left(\frac{ a_2}{f_2}\right) \right),
	\label{Va22}
\eea
where $m_a(T)$ is given by
\bea
m_a(T) &=&{\rm min}\left[ m_a  \lrfp{T}{\Lambda_2}{p},  m_a  \right].
	\label{ma}
\eea
Here we have defined $m_a \equiv \Lambda_2^2/f_2$ 
and $n_1$ ($n_2$) and $f_1$ ($f_2$) are, respectively, the domain wall number and the decay constant for $a_1$ ($a_2$).
$p$ is a negative constant, which depends on the details of the interactions responsible for $V_2$.
In the following numerical calculations we assume the radiation dominated Universe and set 
$p=-3$ for simplicity, but its precise value  is not relevant  for our main results.  
Note that the two axions have a mixing through $V_1$, and so,
 $m_a(T)$ is not necessarily equal to one of the mass eigenvalues.

Let us define two effective decay constants for later use. When the curvature of $V_1$ is larger than that of $V_2$,
one can define  the effective decay constants $f$ and $F$ along the valley of $V_1$ and the direction orthogonal to it,
respectively.  
Specifically,  the lighter direction is determined by $n_1a_1/f_1 + n_2 a_2/f_2 = {\rm const.}$.
The effective decay constants are given by
\begin{align}
f &= \frac{\sqrt{n_2^2 f_1^2 + n_1^2 f_2^2}}{n_1},\\[1mm]
F &= \frac{f_1 f_2}{\sqrt{n_1^2 f_2^2 + n_2^2 f_1^2}}
\end{align}
and one can see that, for large $n_1$ and/or $n_2$, there is a hierarchy between $f$ and $F$.
Such hierarchical decay constant attracted attention in a context of the so called alignment 
mechanism for large-field inflation~\cite{Kim:2004rp, Ben-Dayan:2014zsa,Tye:2014tja}. 
The reason for this choice of the axion potential will be clear shortly. 

\begin{figure}[tp]
\centering
\includegraphics [width = 8cm, clip]{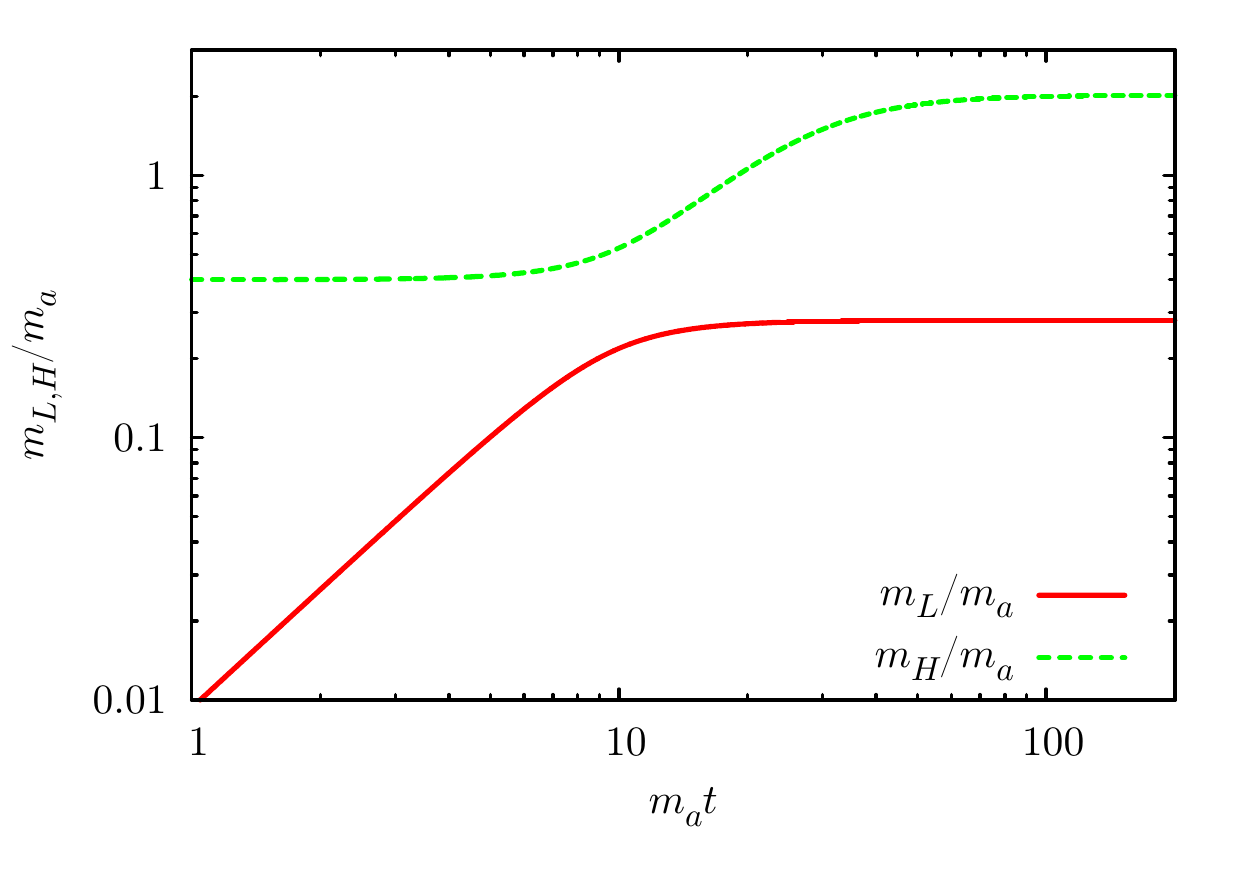}
\caption{
	The time evolution of the mass eigenvalues of two axions, $m_H$ and $m_L$.
}
\label{fig:level_crossing}
\end{figure}

We assume that, while the potential height of $V_1$ is constant with time, 
the potential height of $V_2$ grows with time. 
This is the case e.g., if the gauge sector responsible for $V_2$ is heated to
high temperature where the gauge interactions are weak. Then, at high temperature,
there is a massless flat direction along $n_1a_1/f_1 + n_2 a_2/f_2 = {\rm const.}$.
As the temperature decreases, $m_a(T)$ turns on and
the height of $V_2$ increases. We assume that the mass of $a_2$ finally becomes heavier than $a_1$, and 
then it becomes constant in the low energy. At low temperatures, the heavier mass eigenstate is almost $a_2$, while
the lighter one is $a_1$. See Fig.~\ref{fig:level_crossing} where we show typical evolution of the two mass eigenvalues, $(m_L, m_H)$
with $m_L < m_H$,  while the height of $V_2$ increases due to non-perturbative effects. Here the mass eigenvalues are
evaluated at the potential minimum. Here and in what follows we adopt 
$n_1 = n_2 =10$, $f_1 = f_2 = 10^{15}~\mathrm{GeV}$ and $(\Lambda_1/\Lambda_2)^2 = 0.02$ as reference values.
The level crossing occurs at $m_a t \simeq 10$ when the two mass eigenvalues become
comparable.

The axion potential changes significantly during the level crossing. This might affect the axion dynamics 
if the axion has already started to oscillate before or around the level crossing. 
 If the axion starts oscillations
along the valley of $V_1$  much before the level crossing, the change of the axion potential is so slow compared to the
typical oscillation period that the axion number density in the comoving volume becomes the adiabatic invariant. Then, 
resonant transition takes place during the level crossing  {\it a la} the MSW effect~\cite{Wolfenstein:1977ue,Mikheev:1986gs}. As a result, the final axion density is suppressed by the mass ratio, $m_L/m_H$,  compared to the case without the resonant transition~\cite{Kitajima:2014xla}. 
The adiabaticity is violated if the oscillation amplitude is so large that the anharmonic effect
is relevant. In this case, the resonant transition is not complete and both heavy and light
axions are generated. Then, isocurvature perturbations can be suppressed
for certain parameters, as the produced heavy axions partially cancel the original fluctuations~\cite{Kitajima:2014xla}.

\paragraph{Axion roulette and domain wall formation}
-- Now let us consider the case in which the axion starts to oscillate around or slightly before the time
of the level crossing. 
Then,
the scalar potential significantly changes
even during the first few oscillations of the axion. 
As a result, the direction of oscillations is changed each time the axion oscillates. This is because
the axion effectively gets  kicked into different directions around the turning points of oscillations,
and this lasts until the level crossing is completed.
Note that there is no adiabatic invariant in this process because the motion is not periodic at all.

We have assumed that $V_2$ eventually dominates over the scalar potential. 
Namely, at low temperatures $T \ll \Lambda_2$, the heavier mass eigenstate is approximately 
given by $a_2$, while the lighter one is $a_1$. Therefore,
after the commencement of oscillations, the axion dynamics is likely confined in one of 
the valleys of $V_2$, e.g., around $a_2 \approx 0$. On the other hand, it is not certain if the axion is quickly trapped
in one of the minima of $V_1$. In fact,  if the initial axion energy is sufficiently large, the axion will start to 
run along one of the valleys of $V_2$ soon after the onset of oscillation, passing through 
crests and troughs of $V_1$, until it gets trapped in one of the potential minima.
 
\begin{figure}[tp]
\centering
\includegraphics [width = 8cm, clip]{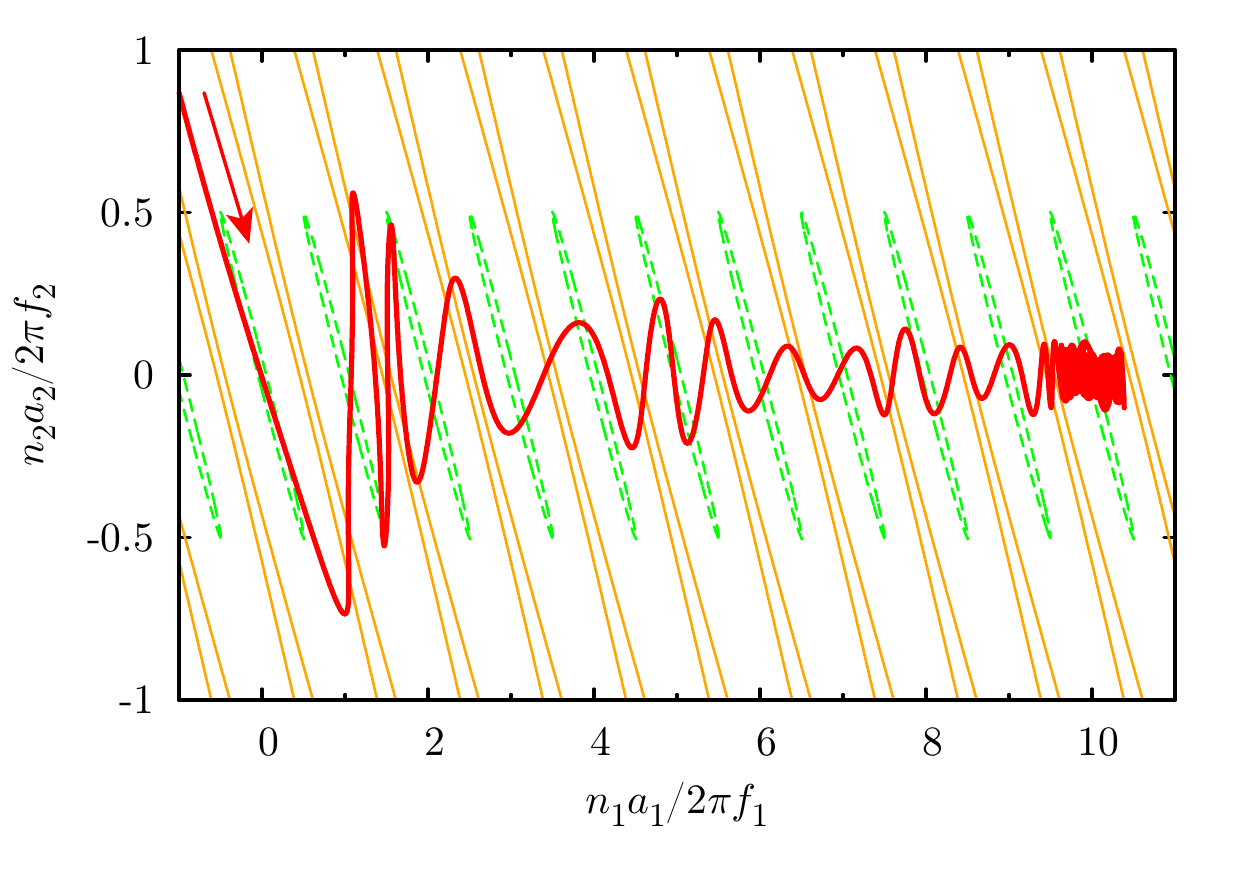}
\caption{
	The trajectory of two axions in the  $a_1$--$a_2$ plane. The axion evolves from left to right.
	The region closed by dashed-green (solid-orange) contour contains the potential minimum (maximum) at $m_a(T)^2f_2^2 = \Lambda_1^4$.
}
\label{fig:trajectory}
\end{figure}

We have numerically confirmed such a behavior.  In Fig.~\ref{fig:trajectory} we show the trajectory of the  two 
axions in the $a_1$--$a_2$ plane. We
adopt the initial condition $a_{2,i}/f_2 = 1.5$ and $n_1 a_{1,i}/f_1 + n_2 a_{2,i}/f_2 = 0$ which minimizes
the $V_1$, but this does not affect our main results.
One can see that the axion first starts to evolve along the flat direction of $V_1$, and
then, gradually starts to run toward positive values of $a_1$ along the valley of $V_2$ (i.e., $a_2=0$). In this example, the axion 
is stabilized after passing through ten crests of $V_1$ shown by the solid-orange lines.
We call the run-away behavior of the axion dynamics the ``{\it axion roulette}."

The axion roulette takes place even if
the axion mass is varied by about one order of magnitude, as long as
the axion starts to oscillate slightly before or around the level crossing. 
According to our numerical calculations, the axion roulette works well for $H_{\rm lc}/H_{\rm osc} = \mathcal{O}(0.1-1)$, 
where the subscripts `lc' and 'osc' imply that the variable is evaluated at the level crossing and at the commencement of oscillations, respectively.
On the other hand,
if the axion starts to oscillate well after the level crossing, the axion initially starts to oscillate
along $a_2$, and no motion along $a_1$ is induced, suppressing the
chaotic run-away behavior.

In order to understand the condition for the axion roulette to take place,
 let us estimate the energy density of the axion $\rho_{a,{\rm osc}}$ at the commencement of oscillations.
In order to  climb over the potential barrier of $V_1$, the initial oscillation energy must be greater than 
the height of $V_1$:
\begin{align} \label{rho_a_osc}
\rho_{a,{\rm osc}} \sim m_{L,{\rm osc}}^2 f^2 \gtrsim  \Lambda_1^4 \sim m_{H,{\rm osc}}^2 F^2.
\end{align}
In deriving the above condition (\ref{rho_a_osc}) we have assumed that the axion starts to move along the valley 
of $V_1$, because the curvature is still dominated by $V_1$ until the level crossing, even if the potential height of $V_2$ 
is larger than $V_1$. One can imagine a cosine potential $V_2$ with grooves induced by $V_1$.
Around the level crossing, the mass eigenvalues at the potential minima are comparable to each other.
Therefore, if the effective decay constant along the flat direction is enhanced,i.e., $f \gg F$,  the initial axion energy easily exceeds
the height of the potential barrier.  This is the reason why we have chosen the axion potential (\ref{Va2}) and (\ref{Va22}),  
where the alignment mechanism is operative. 
Specifically, if one sets $f_1 = f_2$ and $n_1 = n_2 = N$, Eq.~(\ref{rho_a_osc}) yields 
$N \gtrsim {m_H/m_L}|_{\rm osc}$.
Equivalently, one may introduce a hierarchy in the two decay constants from the beginning. 
In contrast to the application to large-field inflation, we do not need
very large hierarchy in the decay constants or super-Planckian ones. The hierarchy by 
a factor of several to ten is sufficient. 

\begin{figure}[tp]
\centering
\includegraphics [width = 8cm, clip]{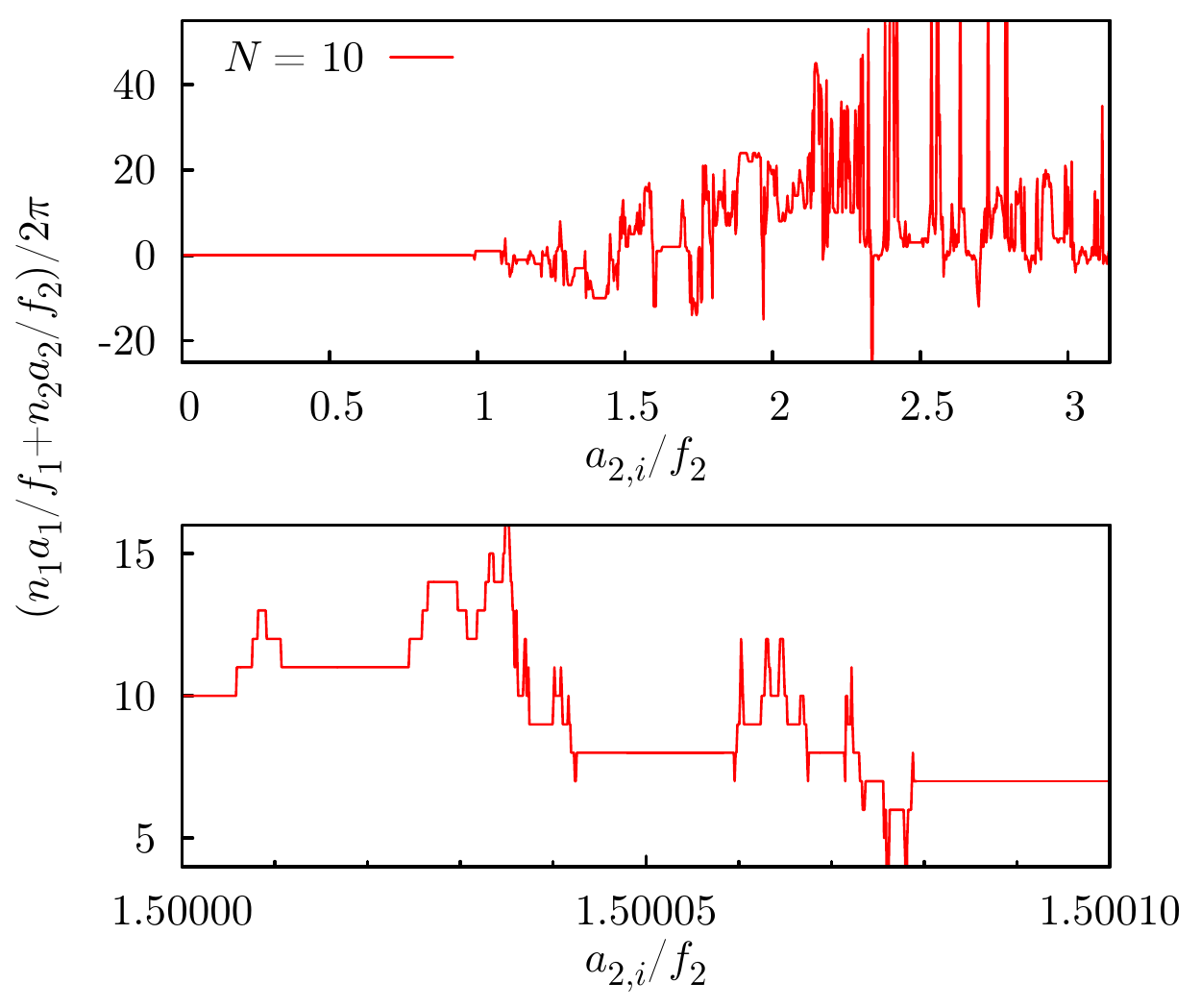}
\caption{
	The final value of $(n_1a_1/f_1+n_2 a_2/ f_2)/2\pi$ as a function of the initial misalignment angle of $a_2$.
	The bottom panel is the zoom-up of the top panel between $a_{2,i}/f_2 = 1.5$~--~$1.5+10^{-4}$.
}
\label{fig:theta1f}
\end{figure}

So far we have studied the spatially homogeneous evolution of the axion, neglecting 
spatial inhomogeneities. The axions acquire quantum fluctuations of order $H_{\rm inf}/2\pi$
during inflation, if they are lighter than the Hubble parameter, $H_{\rm inf}$.
If the final minimum where the axion is stabilized is highly sensitive to the initial position,
the domain walls are likely produced. 
 In Fig.~\ref{fig:theta1f}  we show the final value of $(n_1a_1/f_1+n_2a_2/f_2)/2\pi$ 
 as a function of the initial misalignment 
 angle of $a_2$. 
 The chaotic behavior becomes strong at large misalignment angles because the initial
 oscillation energy is increased. For larger $N$, the chaotic behavior is observed at
 smaller misalignment angles.
Therefore the axion can be stabilized
in different potential minima in each Hubble patch, even if the quantum fluctuation is much smaller
than the decay constant.  In addition, even at subhorizon scales,
spatial inhomogeneities are likely to grow during the time when the axion runs through
crests and troughs of $V_1$. Therefore, the axion roulette is most
probably accompanied by domain wall formation.

\paragraph{Discussion}
-- The domain walls are cosmologically problematic,
and they must be either inflated away, or unstable and decay quickly before dominating the
Universe. The latter can be realized if one introduces another small shift symmetry breaking term
to lift the degeneracy of the potential minima. The domain wall collapse and the emitted gravitational 
waves have been studied extensively in the literature. \cite{Gleiser:1998na,Hiramatsu:2010yz,Kawasaki:2011vv,Hiramatsu:2013qaa}
Our analysis can be straightforwardly applied to the mixing between the QCD and hidden axions.
In this case, the peak frequency of the gravitational waves is comparable to or lower than ${\cal O}(10^{-7})$\,Hz, which may be
within the sensitivity reach of the pulsar timing experiments.\cite{Kramer:2004rwa,IPTA:2013lea}

So far we have introduced a large domain wall number $n_1 = n_2 = N >1$ to realize the required hierarchy in the
effective decay constants. 
Instead, we may introduce hierarchy in the decay constants 
from the very beginning, because the axion dynamics remains unchanged. The property of the domain walls,
however, may be crucially modified. This is the case if the domain wall number $N$ is equal to unity.
In this case, all the potential minima of $V_1$ along the potential valley ($a_2 = 0 $) are identical, and therefore,
one cannot generate any bias between them. If domain walls were bounded by strings, they would be
unstable and collapse quickly after formation. In the case of the axion roulette, no strings are formed,
and so, the domain walls with $N=1$ are stable in a cosmological time scale~\cite{Preskill:1991kd}.  
This is nothing but a cosmological disaster, and the domain walls must be inflated away. The value of $N$ is crucial 
for the fate of domain walls.

The domain wall formation
implies that the primordial quantum fluctuations of the axion result in fluctuations of
order unity through the chaotic dynamics during the level crossing. Therefore, at the domain
wall formation, the initial energy densities of domain walls as well as coherent oscillations
have fluctuations of order unity at superhorizon scales. As the Universe expands, the domain wall network
is known to quickly follow the scaling regime \cite{Press:1989yh,Oliveira:2004he,Leite:2011sc}
Therefore, the energy density of the domain walls in the scaling regime have only fluctuations of order 
unity at subhorizon scales, and their fluctuations at superhorizon scales are suppressed. On the other hand, the axion coherent 
oscillations produced during the domain wall formation remain to have density perturbations 
of order unity at superhorizon scales.  Therefore, those axions should decay before their energy density exceeds
about $10^{-6}$ of the total energy density, since otherwise the non-Gaussianity will be too large.
If the axion domain walls are responsible for the origin of the
baryon asymmetry, isocurvature perturbations and their non-Gaussianity are also generated~\cite{Daido:2015gqa}.
In another case, the axion produced by the domain wall decays may contribute to the cold dark matter.

If the axion roulette works, the kinetic energy of the axion exceeds the height of the potential barrier,
which is impossible to realize in the case of a single axion field. Such a large velocity of the axion
could be used to enhance the baryon asymmetry in the spontaneous baryogenesis scenario~\cite{Cohen:1987vi,Chiba:2003vp}.
We leave a detailed analysis of this scenario, especially the estimate of the isocurvature perturbations
for future work.

In summary, the axion roulette works and the axion domain walls are formed if (i) the axion starts to oscillate 
around or slightly before the level crossing, and (ii) the initial oscillation energy is
sufficiently large to climb over the potential barrier (cf. (\ref{rho_a_osc})).
The first condition does not require any severe fine-tuning:
for instance, this is naturally satisfied if the final two axion masses are comparable 
and the decay constants $f_1$ and $f_2$ are around the GUT scale or higher.
The second condition is met if the alignment mechanism is operative, and it is known
that such enhancement occurs frequently for several axions~\cite{Choi:2014rja,Higaki:2014pja}.
Therefore, the domain wall formation may be a more common phenomenon in the
Axiverse than previously thought.

\acknowledgments
This work was supported by  JSPS Grant-in-Aid for
Young Scientists (B) (No.24740135 [FT]), 
Scientific Research (A) (No.26247042 [FT]), Scientific Research (B) (No.26287039 [FT]), and
 the Grant-in-Aid for Scientific Research on Innovative Areas (No.23104008 [NK, FT]). 
This work was also supported by World Premier International Center Initiative (WPI Program), 
MEXT, Japan [FT].

\bibliography{dwaxiverse}
\end{document}